\documentclass[reprint,superscriptaddress, pra]{revtex4-1}
\pdfoutput=1

\usepackage[utf8]{inputenc}
\usepackage{xspace}
\usepackage[T1]{fontenc}
\usepackage{graphics,graphicx,keyval,subfigure}
\usepackage[normalem]{ulem}
\usepackage{fancyhdr}
\usepackage{stmaryrd}
\usepackage[x11names,svgnames]{xcolor}
\usepackage{amsmath, stackrel, mathtools}
\usepackage{amsfonts,bm,bbm}
\usepackage{upgreek}
\usepackage{etoolbox}
\usepackage{calc}
\usepackage{datetime}
\usepackage[squaren,Gray]{SIunits}
\usepackage{cancel}
\usepackage{amsmath,amsfonts}
\usepackage{amssymb}
\usepackage{mathtools}
\usepackage{slantsc}
\usepackage[hidelinks]{hyperref}
\usepackage[all]{hypcap}
\usepackage{tikz}
\usepackage{lipsum}
\usepackage{blindtext}
\usepackage{perpage}
\usepackage{physics}
\usepackage{textcomp}
\usepackage{bbold}
\usepackage{multirow}
\usepackage{todonotes}
\usepackage[french,english]{babel}
\usepackage{dsfont}

\begin{document}

\title{Error-correcting entanglement swapping using a practical logical photon encoding}
\author{Paul Hilaire}
\affiliation{Department of Physics, Virginia Tech, Blacksburg, Virginia 24061, USA}
\author{Edwin Barnes}
\affiliation{Department of Physics, Virginia Tech, Blacksburg, Virginia 24061, USA}
\author{Sophia E. Economou}
\affiliation{Department of Physics, Virginia Tech, Blacksburg, Virginia 24061, USA}
\author{Frédéric Grosshans}
\affiliation{Sorbonne Université, CNRS, LIP6, F-75005 Paris, France}

\begin{abstract}
Several emerging quantum technologies, including quantum networks, modular and fusion-based quantum computing, rely crucially on the ability to perform photonic Bell state measurements. Therefore, photon losses and the 50\% success probablity upper bound of Bell state measurements pose a critical limitation to photonic quantum technologies.
Here, we develop protocols that overcome these two key challenges through logical encoding of photonic qubits. Our approach uses a tree graph state logical encoding, which can be produced deterministically with a few quantum emitters, and achieves near-deterministic logical photonic Bell state measurements while also protecting against errors including photon losses, with a record loss-tolerance threshold.
\end{abstract}
\maketitle

\section{Introduction}

Photons play a unique role in quantum information technologies. They are the only qubits that can travel over long distances, making them central to applications such as quantum networks and the envisioned quantum internet~\cite{Wehner2018}. The basic building blocks of quantum networks are quantum repeaters~\cite{Briegel1998, Dur1999, Sangouard2011,  Muralidharan2016}, nodes designed to overcome the challenge of photon loss. Repeaters rely on Bell state measurements (BSM) between photons to extend entanglement through the network in a process known as entanglement swapping~\cite{Sangouard2011, Muralidharan2016}.
This in turn enables quantum cryptographic~\cite{Bennett1992b}, computing~\cite{Nickerson2014}, and sensing~\cite{Gottesman2012} protocols.

For distributed quantum computing~\cite{Lim2005, Beige2007}, photonic BSMs are also critical to compose a large-scale quantum computer~\cite{Awschalom2021, Jiang2007b} from small modules of networked matter-based quantum processor, by enabling inter-module entangling gates, through gate teleportation~\cite{Eisert2000}.

Finally, BSMs are the key primitive of a newly introduced model for silicon-photonic quantum information processing, fusion-based quantum computing~\cite{Bartolucci2021}.  This can be thought of as a temporal analog of cluster-state quantum computing, with the advantage that the required resource states are much more modest (constant in the size of the computation). During the computation, pairs of photons coming from different resource states are Bell measured.

While photonic BSM plays an essential role in all these technologies, unfortunately, there is a fundamental limitation in its success probability: it succeeds 50\% of the time~\cite{Weinfurter1994, Braunstein1995, Michler1996, Lutkenhaus1999, Vaidman1999, Calsamiglia2001}.
This is a major obstacle in realizing quantum networks, distributed quantum computing, and photonic quantum computing.
An additional challenge with photonic quantum technologies is photon loss.
Photons can be absorbed, leading to an irreversible loss of the information they encode.
This is particularly problematic for long-distance quantum networks.
These two shortcomings, together severely impede photonic quantum information processing, and often lead to proposals with daunting resource overheads.

Prior work has proposed ways to boost the BSM probability through the use of ancillary photons~\cite{Grice2011, Ewert2014,  Wein2016, Olivo2018}, non-linear interaction with an atom~\cite{Lloyd2001, Kim2001, Kim2002}, or hyperentanglement~\cite{Kwiat1998, Walborn2003, Schuck2006, Barbieri2007} to achieve (near) deterministic photonic BSMs. Nonetheless, these solutions are not tolerant to photon losses and errors. Error reduction requires either photon purification~\cite{Dur1999, Pan2001} or the logical encoding of a qubit on many photons~\cite{Jiang2009}.

In the context of quantum repeater (QR) protocols, error correction is used by third-generation repeaters for loss and error tolerance~\cite{Muralidharan2016}. 
Any quantum error correcting code could in principle be used~\cite{Terhal2015} but they usually require hundreds of matter qubits at each node and efficient light-matter interactions to transfer quantum states between photonic qubits and matter qubits in an efficient way, and these resources have not yet been experimentally demonstrated~\cite{Hacker2016}. All-photonic QRs~\cite{Azuma2015, Pant2017, Ewert2016, Ewert2017} either remove these requirements entirely in the case where they are generated using linear optics, or significantly reduce them when they are produced using a deterministic approach based on a few matter qubits~\cite{Lindner2009, Buterakos2017}.
A logical photonic BSM has been proposed in  Refs.~\cite{Ewert2016, Ewert2017, Lee2019}, initially using a quantum parity code~\cite{Ralph2005} and subsequently extended to arbitrary Calderbank-Shor-Steane codes~\cite{Schmidt2019}. However, large highly entangled states of photons that serve as error-correcting codes are generally difficult to produce~\cite{Bell2014}, largely for the same reason that makes photonic BSMs probabilistic.

In this paper, we simultaneously address both challenges, the probabilistic nature of BSMs and photon loss, through two protocols, which we call ``static'' and ``dynamic'', that allow error-corrected photonic BSMs on logical qubits, with a record loss-tolerance threshold for the dynamic protocol.
We use a tree graph state logical encoding~\cite{Varnava2006, Varnava2007}, which can be generated deterministically with a few matter qubits~\cite{Buterakos2017, Zhan2020}. These generation procedures build on an experimentally-demonstrated protocol~\cite{Lindner2009, Schwartz2016} to produce linear cluster states.

\begin{figure}
  \centering
  \includegraphics[width = 8cm]{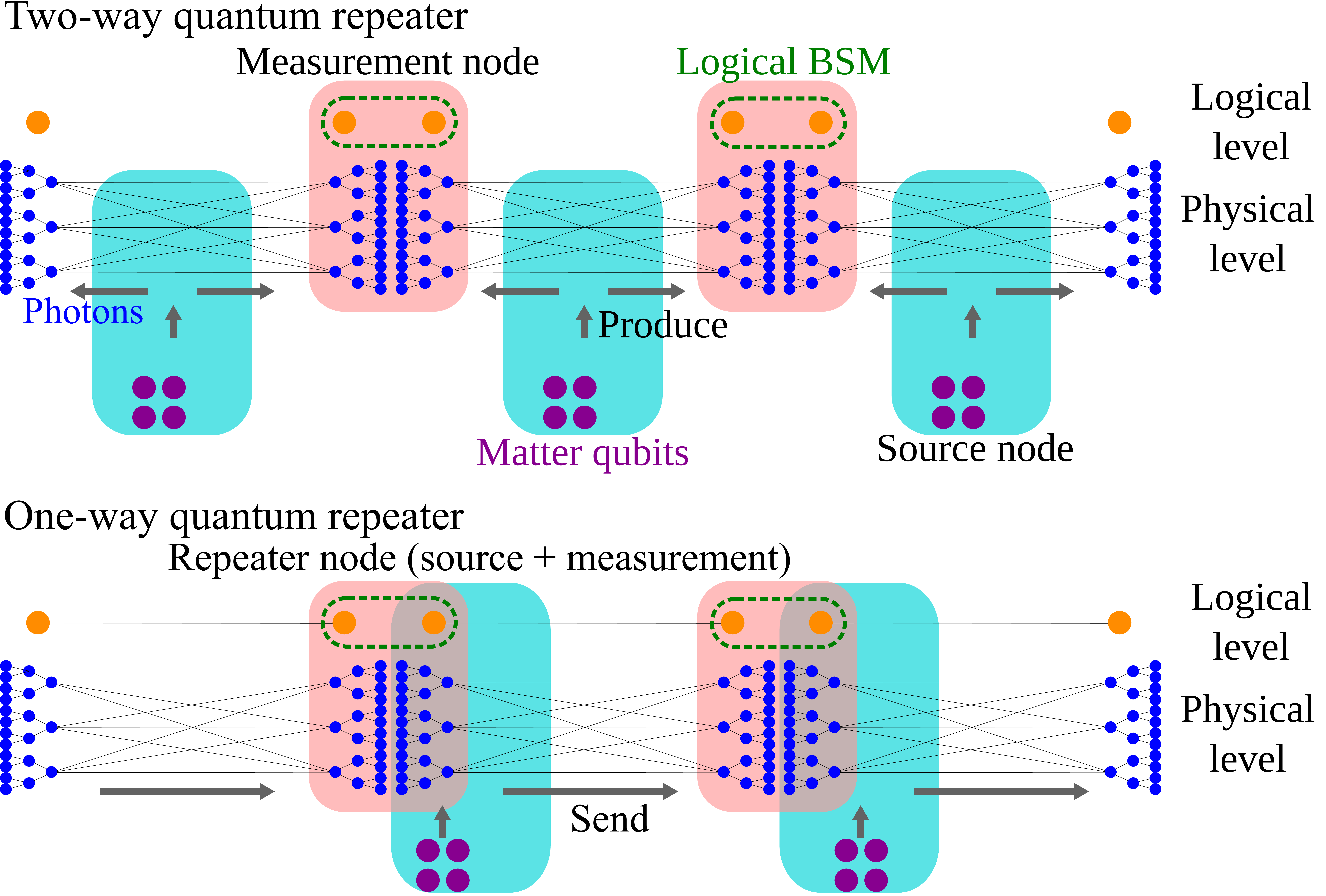}
  \caption{Two-way  (top) and one-way (bottom) QR protocol using deterministic tree graph state generation~\cite{Buterakos2017, Hilaire2020, Zhan2020} with a few matter qubits and the logical BSM protocols studied in this paper (see Appendix~\ref{app_generation} for the deterministic generation procedure).
  }
  \label{fig_QR}
\end{figure}

Our results directly apply to two-way and one-way all-photonic QRs~\cite{Muralidharan2016},
making the original proposal~\cite{Azuma2015} both resource-efficient and fully error-correctable.
In the two-way QR scheme displayed in Fig.~\ref{fig_QR}, a logical Bell pair is produced with a few matter qubits at each source node using the generation sequence detailed in Appendix~\ref{app_generation}, and each logical qubit is sent to an adjacent measurement node where a logical BSM is performed. In the one-way QR scheme, the source nodes and measurement nodes are the same, and one logical qubit is Bell measured at this node while the other one is sent to the next node.

In the remainder of this paper, we describe how to perform a logical BSM, which is the cornerstone, not only of these QR schemes, but also of fusion-based quantum computing.
The paper is organized as follows. In Sec~\ref{sec_BSM}, we introduce the stabilizer formalism to describe two-photon BSMs using linear optics. In Sec.~\ref{sec_logical}, we introduce the logical encoding that we are using. In Sec.~\ref{sec_counterfactual}, we show how to use this encoding to produce logical BSM protocols in a measurement-based setting. Finally, in Sec.~\ref{sec_perf}, we evaluate the performances of the two protocols introduced in this paper.

\section{Two-photon Bell state measurement}
\label{sec_BSM}
A BSM is a joint measurement of two qubits, $a$ and $b$, in one of the four Bell states:
\begin{equation}
  \begin{aligned}
    \ket{\Phi_{ab}^{\pm}} = \frac{1}{\sqrt{2}}\left( \ket{0_a 0_b} \pm \ket{1_a 1_b} \right) \Leftrightarrow \left\{\substack{ \langle Z_a Z_b \rangle =+1\\ \langle X_a X_b \rangle = \pm 1} \right., \\
    \ket{\Psi_{ab}^{\pm}} = \frac{1}{\sqrt{2}}\left( \ket{0_a 1_b} \pm \ket{1_a 0_b} \right) \Leftrightarrow \left\{\substack{ \langle Z_a Z_b \rangle =-1\\ \langle X_a X_b \rangle = \pm 1} \right..
  \end{aligned}
  \label{eq_two_photon_bsm}
\end{equation}
As shown on the right-hand side of Eq.~\eqref{eq_two_photon_bsm}, using the stabilizer formalism~\cite{Gottesman1997}, a BSM can also be interpreted as the measurement of the two operators $X_a X_b$ and $Z_a Z_b$, where $X$, $Y$, and $Z$ are the usual Pauli matrices, and the subscript indicates on which qubit the operator is applied.
Experimentally, a linear optical BSM~\cite{Weinfurter1994, Braunstein1995, Michler1996} measures one of these operators, and, depending on the outcome of this measurement, $+1$ or $-1$, the second operator is measured or not.
In the following, we consider a setup where we can discriminate $Z_a Z_b$ unambiguously and $X_a X_b$ half of the time (when $Z_a Z_b$ has parity $ 1$), even though
all configurations of operators and measurement outcomes are experimentally feasible.

Hence, denoting by $\eta$ the detection probability of each photon, a two-photon linear optical BSM can yield three different results:
a \emph{complete} measurement (with probability $\eta^2 / 2$), i.e.\ $X_a X_b$ and $Z_a Z_b$ are measured; a \emph{partial} measurement (with probability $\eta^2 / 2$), i.e.\ only $Z_a Z_b$ is measured; or a \emph{failed} measurement (with probability $1 - \eta^2$), i.e.\ no outcome is measured, if at least one photon is lost.

\begin{figure}
  \centering
  \includegraphics[width = 8cm]{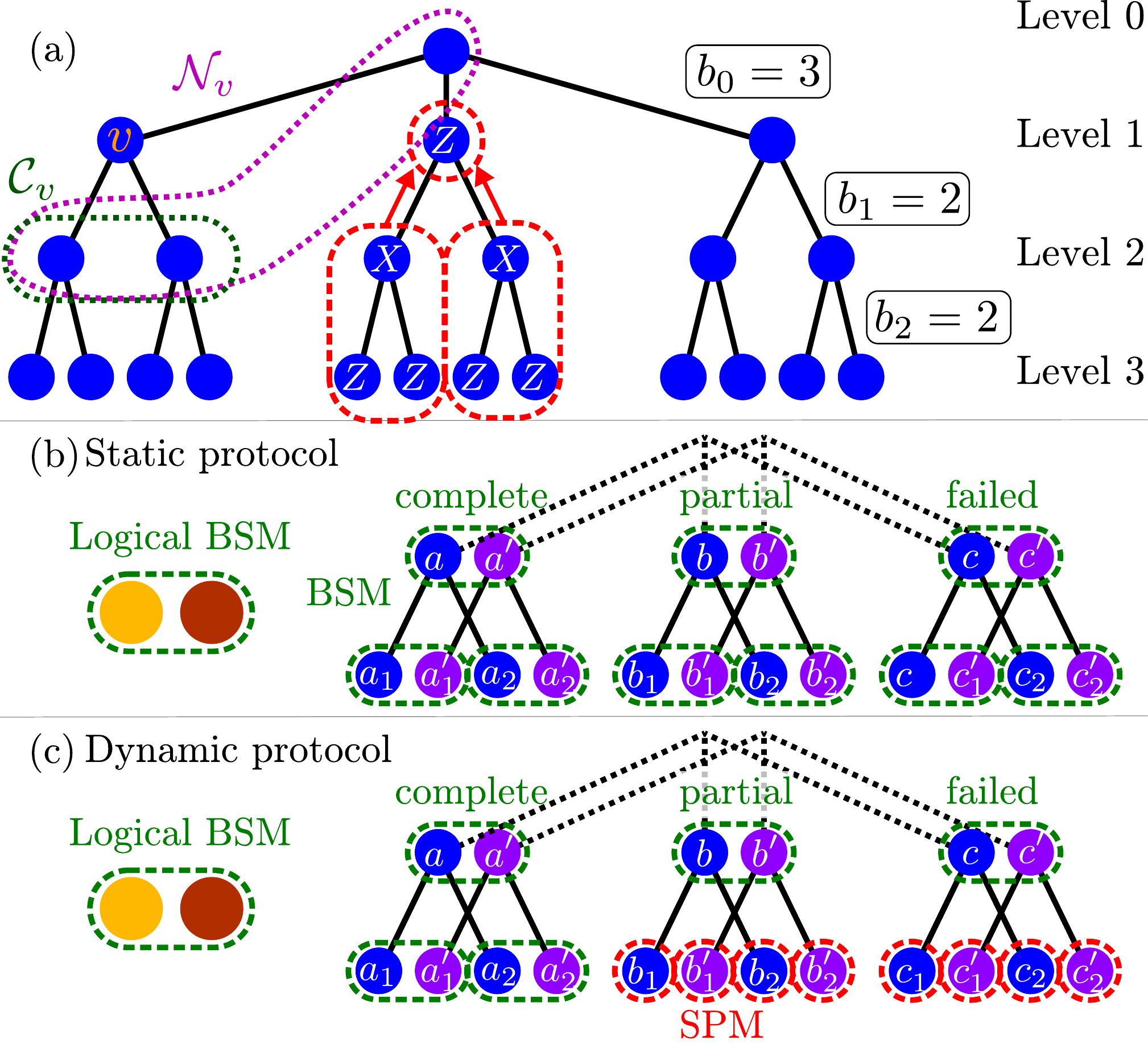}
  \caption{(a) Tree graph state and notations. $\mathcal{N}_v$ and $\mathcal{C}_v$ denote the set of neighbor qubits and child qubits of vertex $v$. Indirect $Z$ measurements are also illustrated. (b, c) Logical BSM at the logical level (left panel) and at the physical level (right panel) for the static (b) and the dynamic (c) protocols. (SPM: single-photon measurements.)}
  \label{fig_tree_bsm}
\end{figure}

\section{Logical Bell state measurements}
\label{sec_logical}
To avoid this limitation and to enable loss-tolerance and error reduction, we are using a tree graph state encoding which is a stabilizer error-correcting code.
A graph state $\ket{G}$ is the unique quantum state described by a graph $G=(V,E)$, with a set of vertices $V$ corresponding to qubits, and edges $E$, which is stabilized by the $|V|$ stabilizers $K_v$ for $v \in V$~\cite{Hein2004} :

\begin{equation}
  K_v \ket{G} = \left(X_v \smashoperator{\prod_{w \in \mathcal{N}_v}} Z_w \right) \ket{G} = \ket{G},
  \label{eq_stabilizer}
\end{equation}
where $\mathcal{N}_v = \{w| (v,w) \in E\}$ is the set of qubits neighboring qubit $v$ (see Fig~\ref{fig_tree_bsm}(a)).

More specifically, we are interested in tree graph states of depth $d$ that are defined by a branching vector $\vec{b}=(b_0, b_1, ..., b_{d-1})$,
which can  encode a logical qubit~\cite{Varnava2006} (see Appendix~\ref{app_tree} for more details).
In this encoding, the physical $X$ and $Z$ operators are replaced by logical operators $X_L$ and $Z_L$:
\begin{equation}
  \begin{aligned}
    X_L & = X_v  \prod_{w \in \mathcal{C}_v} Z_w, \\
    Z_L & = \prod_{u \in \mathcal{C}_0} Z_u,
  \end{aligned}
  \label{eq_logical_sqm}
\end{equation}
where $v$ is any qubit from level 1 of the tree ($v\in\mathcal{C}_0$), and
$\mathcal{C}_v$ denotes the set of child qubits of $v$ (i.e. \ the qubits from level 2 that are neighbors of $v$).

It directly follows from Eqs.~\eqref{eq_two_photon_bsm} and~\eqref{eq_logical_sqm} that a complete BSM on two logical qubits encoded with trees requires the measurement of both $X^{ }_L X_L'$ and $Z^{ }_L Z_L'$ (we use a prime to denote the second tree):
\begin{equation}
  \begin{aligned}
    X^{ }_L X_L' & = (X_v X_{v'}) {\hskip -0.7cm} \prod_{(w,w') \in (\mathcal{C}_v, \mathcal{C}_{v'})} {\hskip -0.7cm} (Z_w Z_{w'}),  \\
    Z^{ }_L Z_L' & =  {\hskip -0.7cm} \prod_{(v,v') \in (\mathcal{C}_0, \mathcal{C}_{0'})}  {\hskip -0.7cm} (Z_v Z_{v'}),
  \end{aligned}
  \label{eq_bsm_l}
\end{equation}
where $v\in\mathcal{C}_0$ is any qubit from the first level of one tree, and $v'\in\mathcal{C}_{0'}$ is its counterpart from the other tree. In this expression, we have paired the operators $X_v X_{v'}$ and $Z_v Z_{v'}$ to highlight that these logical measurements can be implemented by physical two-photon BSMs that combine each photon from one tree with its counterpart from the second tree.

In Fig.~\ref{fig_tree_bsm}(b), we illustrate this strategy, which we call the ``static'' protocol, with two trees with branching vectors $\vec b = (3, 2)$.
If a physical BSM on level 1 qubits, say $a$ and $a'$, is complete (thus yielding a measurement of $X_a X_{a'}$), and if the BSMs on all the child qubits of $a$ and $a'$ are at least partial (thus yielding a measurement for $Z_{a_1}Z_{a'_1}$ and $Z_{a_2}Z_{a'_2}$), we have performed a successful logical $X^{ }_L X'_L$ measurement on the two logical qubits.
A successful logical $Z^{ }_L Z'_L$ measurement would also require all the remaining physical BSMs on the first-level qubits, $b,b'$ and $c,c'$, to be at least partial to yield $Z_b Z_{b'}$ and $Z_c Z_{c'}$.
In the absence of photon losses ($\eta=1$) and errors, all physical BSMs are at least partial, and only one BSM on the level 1 qubits should be complete. This already boosts the overall success probability
to $1-2^{-b_0}$, which can be made arbitrarily close to $1$ by increasing the number $b_0$ of first-level qubits in the tree encoding.

At this point, it is worth clarifying some subtleties related to replacing logical two-qubit measurements with a series of physical BSMs. If the goal is to project the logical qubits onto one of the four logical Bell states, then the static protocol will not work, because the physical measurements provide too much information about the state, collapsing the logical qubits to a separable state. However, for applications such as qubit or gate teleportation and entanglement swapping, which are the main applications of photonic BSMs, this is not an issue. In such applications, one or both logical qubits are initially entangled with additional qubits. Upon success, the static protocol will still generate all the same entanglement links or teleported states on the unmeasured qubits that one would expect from a BSM. This is explained in detail in Appendix~\ref{app_counterfactual}.

\section{Correcting losses and errors in logical Bell state measurements}
\label{sec_counterfactual}
The loss-tolerance and error-correcting properties of this logical BSM naturally arise from the counterfactual error correction properties of the tree graph states that were already demonstrated in Refs.~\cite{Varnava2006,Azuma2015} for single logical qubit measurements. Below, we show that these properties can be extended to logical BSMs.
Before we explain this, we first recall that single-qubit counterfactual error correction is based on indirect measurements of a single qubit $v$ in a tree,  performed by measuring other qubits using the stabilizing properties of the graph.
The loss-tolerance builds on the fact that a measurement can be realized even if the qubit is lost, while error correction is based on multiple indirect measurements of the same single qubit and the use of a majority vote to reduce errors.

From  Eq.~\eqref{eq_stabilizer}, it indeed follows that a qubit $r \in \mathcal{N}_w$ can be indirectly measured in the $Z$ basis by measuring $Z_r K_w = X_w \prod_{s \in \mathcal{N}_w}^{s \neq r} Z_s$, i.e., an $X$ measurement on $w$ and $Z$ measurements on all its neighbors except $r$. It is therefore possible to indirectly $Z$-measure a qubit $r$ of a tree graph state by measuring one of its child qubits $w$ and the qubit set $\mathcal{C}_w$ (see Fig.~\ref{fig_tree_bsm}(a)).

Regarding two-qubit measurements and following the notations in Fig.~\ref{fig_tree_bsm}(b), if the BSM on qubits $c$ and $c'$ fail, we can still recover $Z_c Z_{c'}$ indirectly
using the deeper level qubits,
by measuring either $Z_c  K_{c_1} Z_{c'} K_{c'_1} = X_{c_1} X_{c'_1}$ or  $Z_c  K_{c_2} Z_{c'} K_{c'_2} = X_{c_2} X_{c'_2}$.
In addition, by performing the same measurement indirectly multiple times, we can use a majority vote to allow for error correction.
We introduce here the \emph{static protocol}, where the logical BSM is recovered by performing two-photon BSMs jointly on each qubit from a tree and its counterpart from the second tree. In the following, we calculate the success probability of such a logical BSM and demonstrate its loss tolerance. The error analysis is included in Appendix~\ref{app_error}.

\paragraph*{Static protocol}
We use the notations for a measurement event of the $W$ observable at the tree level $k$:
\begin{itemize}
  \item $\mathcal{D}_{W, k}$ is a direct measurement.
  \item $\mathcal{S}_{W, k}$ is a single indirect measurement event, using only one stabilizer $K_v$, with $v$ one of its child qubits (at level $k+1$), e.g.\@ the indirect measurement of $Z_cZ_c'$ through the measurement of $X_{c_1}X_{c'_1}$ in the previous example.
  \item $\mathcal{I}_{W, k}$ is an indirect measurement event that can be realized on a collection of $\mathcal{S}_{W,k}$, e.g.\@ when we can try either $X_{c_1}X_{c'_1}$  or  $X_{c_2}X_{c'_2}$  to perform the $Z_{c}Z_{c'}$ measurements.
  \item $\mathcal{M}_{W, k}$ is a measurement event (direct or indirect).
\end{itemize}

We denote by $\mathrm{Pr}[A]$ the probability of an event $A$. For example, $\mathrm{Pr}[\mathcal{D}_{X, k}] = \mathrm{Pr}[\mathcal{D}_{Z, k}] = \eta$ for single-qubit measurements and $\mathrm{Pr}[\mathcal{D}_{XX', k}] = \mathrm{Pr}[\mathcal{D}_{ZZ', k}] / 2 = \eta^2 / 2$ for a two-photon BSM.
Since these probabilities do not depend on the level of the photons in the tree, we simplify the notation by not specifying the level $k$ of the photon for a direct measurement event, $\mathcal{D}_{W,k} = \mathcal{D}_{W}$.

In the following, we consider measurements in the $W$ basis at level $k$ for $W=Z$ or $ZZ'$. $W= ZZ'$ is useful for the static protocol, and $W=Z$ is useful not only for the logical single-qubit measurement~\cite{Varnava2006,Azuma2015} but also in the case of the dynamic protocol that will be introduced next.
The success probability of a direct or indirect measurement event is therefore given by
\begin{equation}
    \mathrm{Pr}[\mathcal{M}_{W,k}] = \mathrm{Pr}[\mathcal{D}_{W,k}] + (1 - \mathrm{Pr}[\mathcal{D}_{W,k}]) \mathrm{Pr}[\mathcal{I}_{W,k}],
    \label{eq_pr_m_wk}
\end{equation}
which states that a measurement is successful if its direct measurement $\mathcal{D}_{W,k}$ is successful or, if not, if its indirect measurement $\mathcal{I}_{W,k}$ is successful.

We should also note that the qubits at the last level of the tree, i.e.\ $k=d$, can only be measured directly such that $\mathrm{Pr}[\mathcal{I}_{W,d}] = 0$.

An indirect measurement of a qubit at level $k\neq d$ can in principle be performed $b_k$ times, and only one needs to succeed:
\begin{equation}
    \mathrm{Pr}[\mathcal{I}_{W,k}] = 1 - (1 - \mathrm{Pr}[\mathcal{S}_{W,k}])^{b_k}.
    \label{eq_pr_i_wk}
\end{equation}
$\mathcal{S}_{W,k}$ depends on the measurements that are realized. For a $W=Z$ measurement (or respectively for a $W=ZZ'$ measurement), we should directly measure
$\widetilde{W}=X$ ($\widetilde{W}=XX'$) on
one of its child qubits $v$ at level $k+1$ and measure directly or indirectly $Z$ ($ZZ'$) all the child qubits of $v$ at level $k+2$.
Therefore,
\begin{equation}
    \mathrm{Pr}[\mathcal{S}_{W,k}] = \mathrm{Pr}[\mathcal{D}_{\widetilde{W},k+1}] \mathrm{Pr}[\mathcal{M}_{W,k+2}]^{b_{k+1}}.
    \label{eq_pr_s_wk}
\end{equation}

As we can see, this set of equations, Eqs.~\eqref{eq_pr_m_wk} -- \eqref{eq_pr_s_wk}, is recursive since the measurement probability of $\mathcal{M}_{W,k}$ depends on the probability of $\mathcal{M}_{W,k+2}$. This explains why the success probability can be increased, and as shown in Appendix~\ref{app_error}, the error probability can be reduced.

With this set of equations we can compute success and error probabilities for the static logical BSM protocol. Indeed,
\begin{align}
        \mathrm{Pr}[\mathcal{M}_{X_L X_L'}] & = \mathrm{Pr}[\mathcal{I}_{ZZ', 0}], \\
        \mathrm{Pr}[\mathcal{M}_{Z_L Z_L'}] & = \mathrm{Pr}[\mathcal{M}_{ZZ', 1}]^{b_0}.
\end{align}

However, the probability of realizing a complete logical BSM, denoted $\mathcal{M}_{{\rm BSM}, L}^{(c)}$, is not the product of $\mathrm{Pr}[\mathcal{M}_{X_L X_L'}]$ and $\mathrm{Pr}[\mathcal{M}_{Z_L Z_L'}]$ since there are correlations between these events ($\mathrm{Pr}[\mathcal{M}_{{\rm BSM}, L}^{(c)}] = \mathrm{Pr}[\mathcal{M}_{X_L X_L'} \cap \mathcal{M}_{Z_L Z_L'}] = \mathrm{Pr}[\mathcal{M}_{X_L X_L'} | \mathcal{M}_{Z_L Z_L'}] \mathrm{Pr}[\mathcal{M}_{Z_L Z_L'}]$):
\begin{widetext}
  \begin{align}
      \mathrm{Pr}[\mathcal{M}_{{\rm BSM}, L}^{(c)}]& =
      \smashoperator[r]{\sum_{\substack{m^{(c)} + m^{(p)} + m^{(f)} = b_0 \\m^{(c)} \geq 1 \\m^{(p)}, m^{(f)} \geq 0 }}} P_{\rm BSM}(m^{(c)}, m^{(p)}, m^{(f)}) \mathrm{Pr}[\mathcal{M}_{Z_L Z_L'}|m^{(f)}] \mathrm{Pr}[M_{X_L X_L'}|m^{(c)}],
      \label{eq_pr_m_bsm_l}
      \intertext{where}
      P_{\rm BSM}(m^{(c)}, m^{(p)}, m^{(f)})
      & = \frac{(m^{(c)}+ m^{(p)}+ m^{(f)})!}{ m^{(c)}! m^{(p)}! m^{(f)}!} \left(\frac{\eta^2}{2} \right)^{(m^{(c)} + m^{(p)})}
      \left(1 - \eta^2\right)^{m^{(f)}}
      \label{eq_bsm_mc_mp_n}
\end{align}
\end{widetext}
is the combinatorial probability of having $m^{(c)}$ complete, $m^{(p)}$ partial and  $m^{(f)}$ failed BSM outcomes at the first level,
with
\begin{equation}
    \mathrm{Pr}[\mathcal{M}_{Z_L Z_L'}|m^{(f)}] = \mathrm{Pr}[\mathcal{I}_{ZZ',1}]^{m^{(f)}},
  \label{eq_m_zz_l_mc_mp}
\end{equation}
the probability to completely measure $Z_L Z_L'$ given that $m^{(f)}$ BSMs failed at the first level,
and with
\begin{equation}
  \mathrm{Pr}[\mathcal{M}_{X_L X_L'}|m^{(c)}] =  1 - \left(1 - \mathrm{Pr}[\mathcal{M}_{ZZ',2}]^{b_1}\right)^{m^{(c)}},
    \label{eq_m_xx_l_mc}
\end{equation}
the probability to completely measure $X_L X_L'$ given that $m^{(c)}$ BSMs at the first level were complete.
With this set of equations, it is possible to calculate the performance of the static protocol.

\paragraph*{Dynamic protocol}
If we allow adaptive measurements, i.e.\@, the measurement basis now depends on the outcomes of previous measurements, we can also build an improved ``dynamic'' logical BSM protocol.
We first note that while we can use the child qubits to perform indirect $Z_v$ or $Z_v Z_{v'}$ measurements, it is impossible to use them to perform indirect $X_v$ or $X_v X_{v'}$ measurements, since measurements on the parent qubits would also be needed.
Therefore, instead of using indirect BSMs to achieve complete BSMs, the objective is to upgrade failed BSMs to partial BSMs by indirectly measuring $Z_v Z_{v'}$ via single-qubit measurements on their child qubits as illustrated in Fig.~\ref{fig_tree_bsm}(c). Thus, in the dynamic protocol, BSMs (single-qubit measurements) are performed on child qubits if the BSM on the parents is complete (partial or failed). Taking the example of Fig.~\ref{fig_tree_bsm}(c) where the BSM on qubits $c$ and $c'$ fails, we can replace the indirect measurement of $Z_c Z_{c'}$ by individual indirect measurements $Z_c$ and $Z_{c'}$. These measurements would succeed with higher probability because they can succeed for example even if qubits $c_1$ and $c'_2$ are lost, thus resulting in a better loss-tolerance of the dynamic protocol compared to the static protocol.  We can also show that it performs better in terms of error correction. The mathematical framework of the dynamic protocol is detailed in Appendix~\ref{app_dynamic}.

\begin{figure}
  \centering
  \includegraphics[width = 7cm]{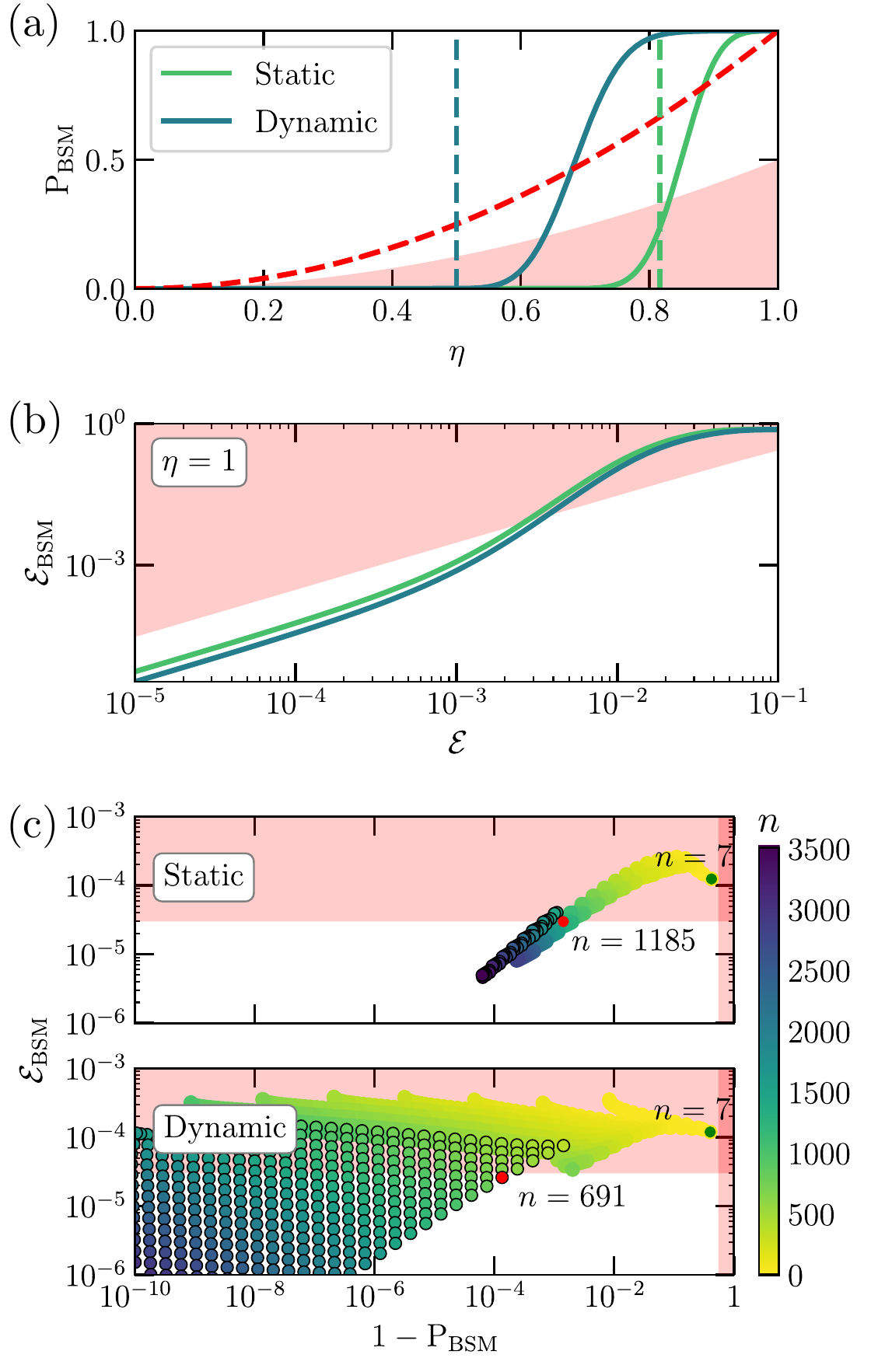}
  \caption{(a) Success probability of a BSM as a function of the single-photon detection efficiency $\eta$ for the static and the dynamic protocols for a tree of branching vector $\vec{b} = (15,15,2)$. Dashed curves correspond to ${\rm P}_{\rm BSM} = \eta^2$ (red), $\eta = 1/2$ (blue), and $\eta = \sqrt{2/3}$ (green). (b) Logical BSM error as a function of the single-qubit depolarization error rate $\mathcal{E}$ for the static and dynamic protocols. (c) Performance of the static (top) and dynamic (bottom) protocols as a function of the number of photons in the tree, for $\eta=95\%$ and $\mathcal{E}=10^{-5}$. Red point: smallest tree which is both error-correcting and loss-tolerant. Green point: smallest tree which is loss-tolerant. Points circled in black (uncircled): depth-3 (depth-2) trees.
   In all these figures, pink regions: no advantage over a two-photon BSM.
   }
  \label{fig_results}
\end{figure}

\section{Performances}
\label{sec_perf}
We now investigate the performance of these protocols.
In Fig.~\ref{fig_results}(a), we evaluate their loss-tolerance using trees with branching vector $\vec{b}=(b_0,b_1,b_2) = (15,15,2)$ (below we show that this tree structure yields good performances  for both loss and error correction).
For high enough single-photon detection probabilities, both protocols perform a near-deterministic logical BSM.
Notably, they overcome the $\eta^2$ limit, which is the upper bound for BSMs with physical qubits, evidencing that they are also loss-tolerant.
Further numerical calculations show that there always exist tree structures that allow an arbitrarily high success probability  as long as $\eta$ is above  $\sqrt{2/3}$ for the static protocol, and $1/2$ for the dynamic protocol.
The dynamic threshold $1/2$ is significantly lower than the generic loss-tolerance threshold $\eta > 1/\sqrt 2$ established in Ref.~\cite{Lee2019} (derived from the bound $\eta \eta' > 1/2$ for symmetric loss $\eta = \eta'$).
This is possible because Ref.~\cite{Lee2019} assumes the logical BSM only uses linear optical BSM, an assumption broken by
our dynamic protocol which also uses single qubit measurements.
This dynamic threshold is in fact the same as for single logical qubit measurements~\cite{Varnava2006} and corresponds to the actual maximum amount of loss that can be corrected with a logical encoding according to the no-cloning theorem.

Considering the same tree and only single qubit depolarization errors, $\mathcal E>0$ but $\eta = 1$,
we show in Fig.~\ref{fig_results}(b) that these protocols are also error-correcting, with a logical BSM error reduced below the rate expected for a linear optical BSM.
As expected, both for loss-tolerance and error-correction, the dynamic protocol outperforms the static protocol.

We now evaluate the performance of these protocols as a function of the number of photons per tree, $n$.
We consider a single-photon detection probability of $\eta = 95 \%$ and an error probability of $\mathcal{E} = 10^{-5}$. Figure~\ref{fig_results}(c) represents the logical BSM success probability and error rate for trees constituted of $n$ photons for the static and the dynamic protocols.
Here, we only present results for trees, found through a systematic numerical search, which have improved performance for either loss or error correction compared to smaller trees. Fig.~\ref{fig_results}(c) shows that the loss-tolerance is easily achieved even with a reduced number of photons; for example only 7 qubits per tree are needed for a photon loss rate of 5\%. However, it takes a tree with $n \geq 1185$ photons (branching vector $\vec{b}=(74, 15)$) to achieve error correction in the static protocol. The dynamic protocol also significantly reduces the amount of resources required since error correction can be achieved with $n \geq 691$ photons using a tree with $\vec{b}=(15, 15, 2)$ branching vector. Notably, a larger number of trees are of interest for the dynamic protocol compared to the static protocol. Our calculations show that encodings using trees of depth-3 are much more tolerant to losses and errors when using the dynamic protocol. Further calculations also show that the size of the tree that achieves error-correction strongly depends on the single-photon detection probability $\eta$.

Regarding the implementation of these logical BSM protocols, we should highlight that, in the static protocol, the photons are always measured via two-photon BSMs. It can therefore be implemented using a static standard linear-optical setup. The dynamic protocol has better performance, but it is also more challenging to implement since photons should be measured in a given order (from the first levels to the deeper levels) and the measurement setting depends on previous detections, thus requiring active components in the optical detection setup that can quickly switch between single-qubit $X$ or $Z$ measurements and the two-photon BSM.

We have shown that it is possible to perform a loss-tolerant and error-corrected logical BSM on photonic qubits encoded with tree graph states, a logical encoding that can be deterministically generated with a few matter qubits. Our results should impact the wide range of quantum technologies that involve photons, including fusion-based quantum computing~\cite{Bartolucci2021}, distributed quantum computing, and quantum networks. The latter application includes a new all-photonic QR protocol that builds on the original proposal of Ref.~\cite{Azuma2015} and requires significantly fewer resources while also enabling error-correction, a feature that was lacking from the original proposal. This QR protocol is based on Bell pairs of logically encoded qubits as shown in Fig.~\ref{fig_QR}. We leave the performance analysis of such new all-photonic QRs to future work.  In addition, as the error correction is limited by the single-photon losses, the performance of such a protocol may drastically increase by using ancilla qubits or non-linear interactions with atoms to further improve the physical qubit BSM success rate. Besides, the transversal nature of the static protocol should allow its generalization to other stabilizer codes, hopefully resulting to more efficient and more robust logical BSMs,
likely at the cost of a more demanding experimental state generation.

\begin{acknowledgments}
  We thank Clément Meignant for stimulating discussions, and Anthony Leverrier, Yuan Zhan, and Shuo Sun for their comments on the manuscript.
  This research was supported by the NSF (Grant No. 1741656) and in part by the EU Horizon 2020 programme (GA 862035 QLUSTER).
  FG acknowledges support  of  the  ANR  through  the  ANR-17-CE24-0035  VanQute project.
\end{acknowledgments}

\appendix

\section{Efficient generation of the logical Bell state}
\label{app_generation}

\begin{figure*}
  \centering
  \includegraphics[width = 12cm]{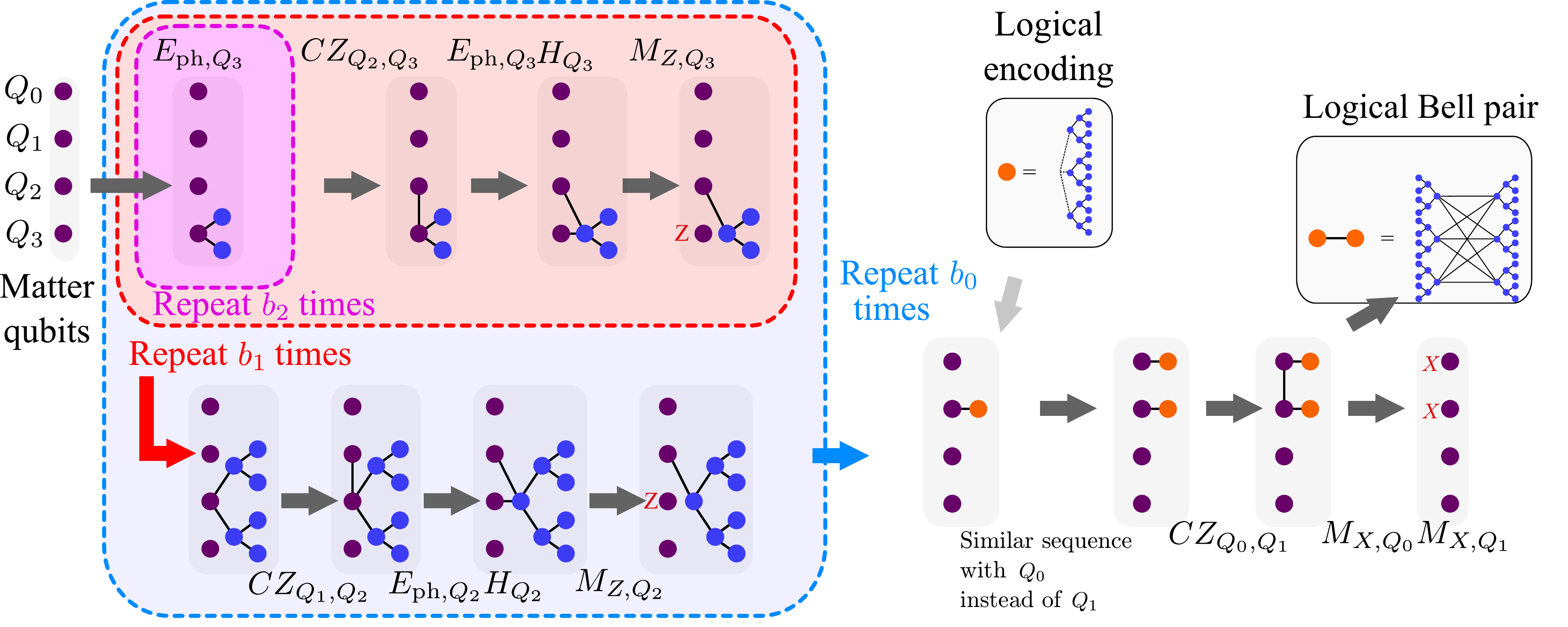}
  \caption{Protocol for the deterministic generation of a logical Bell pair using matter qubits.}
  \label{fig_generation}
\end{figure*}

To operate as a QR, we not only need to be able to perform logical Bell state measurements for the entanglement swapping but we also need to efficiently generate the logical Bell states encoded with tree graph states.
An arbitrary-sized logical Bell-pair can be generated deterministically using a few matter qubits, by using a variant of the generation procedures introduced in Refs.~\cite{Buterakos2017} or in~\cite{Zhan2020}. We illustrate this by adapting the generation procedure of Ref.~\cite{Buterakos2017} to produce a logical Bell pair encoded with tree graph states of depth $d$, following a given sequence based on four operations on matter qubits: the emission of a photon maximally entangled with the matter qubit $E_{\rm ph}$, the Hadamard gate $H$, measurements in the Pauli bases $M_X$, $M_Y$, $M_Z$ and the CZ gate. A logical Bell pair of depth $d$ with branching vector $\vec b = (b_0, b_1, ..., b_{d-1})$ is produced using $d+1$ matter qubits by the following sequence (the operations are applied sequentially from right to left):
\begin{equation}
    M_{X,Q_0} M_{X,Q_1} CZ_{Q_0, Q_1} F(Q_0, \vec b) F(Q_1, \vec b).
\end{equation}
Here, $M_{A, Q_i}$ corresponds to the measurement of qubit $Q_i$ in basis $A$, $CZ_{Q_i, Q_j}$ corresponds to a $CZ$ gate between qubits $Q_i$ and $Q_j$, and the sequence function $F(Q_i, \vec b)$ is defined such that
\begin{equation}
    \begin{aligned}
    F(Q_i, \vec b) & = \left(M_{Z, Q_2} H_{Q_2} E_{{\rm ph, Q_2}} CZ_{Q_i, Q_2} G_2 \right)^{b_0}, \\
    {\rm with \;}G_i & = \left(M_{Z, Q_{i+1}} H_{Q_{i+1}} E_{{\rm ph}, Q_{i+1}} CZ_{Q_{i}, Q_{i+1}} G_{i+1} \right)^{b_{i-1}}, \\
    {\rm and \;}G_d & = \left(E_{{\rm ph, Q_{d+1}}}\right)^{b_{d-1}},
    \end{aligned}
\end{equation}
where we have omitted the single photon rotation for simplicity.
An illustration of this generation sequence is given for $\vec b = (3,2,2)$ in Fig.~\ref{fig_generation}.

\section{Logical encoding with trees}
\label{app_tree}

Here, we explicit what are the logical states $\ket{0_L}$ and  $\ket{1_L}$, using a tree logical encoding, introduced by Ref.~\cite{Varnava2006}.

A tree graph state $\ket{T_{\vec{b}}}$ is defined by a tree with branching vector $\vec{b}$. By using the notations $\vec{b}_i = (b_i, b_{i+1}, ..., b_{d-1})$, with $\vec{b}_0 = \vec{b}$ and $\vec{b}_d = \vec{0}$, we can construct $\ket{T_{\vec{b}}}$ recursively:
\begin{equation}
  \begin{aligned}
    \ket{T_{\vec{b}}} & = \ket{T_{\vec{b}_0}}, \\
    \ket{T_{\vec{b}_i}} & = \ket{0}_i \otimes \ket{T_{\vec{b}_{i+1}}}^{\otimes b_i} + \ket{1}_i \otimes \ket{\bar{T}_{\vec{b}_{i+1}}}^{\otimes b_i},  \\
    \ket{\bar{T}_{\vec{b}_i}} & = Z_i \ket{T_{\vec{b}_i}}, \\
    \ket{T_{\vec{b}_d}} & = \ket{0}_d + \ket{1}_d,\\
  \end{aligned}
  \label{eq_tree}
\end{equation}
where here and in the following, we omit the normalization factors.

The second line of Eq.~\ref{eq_tree} explicits that a tree of depth $i$ with branching vector $\vec{b} = (b_0, ... b_{i-1})$ is composed of a root qubit in state $\ket{+}$ attached to $b_0$ trees of depth $i-1$ with branching vector $\vec{b'} = (b_1, ... b_{i-1})$, with $CZ$ gates.
$\ket{\bar{T}_{\vec{b}_i}}$ is the same state as $\ket{T_{\vec{b}_i}}$ except that a $Z$ operator has been applied on its root qubit $i$. The fourth line ends the recursion to generate a tree of depth $d$.

\begin{figure}
  \centering
  \includegraphics[width = 8cm]{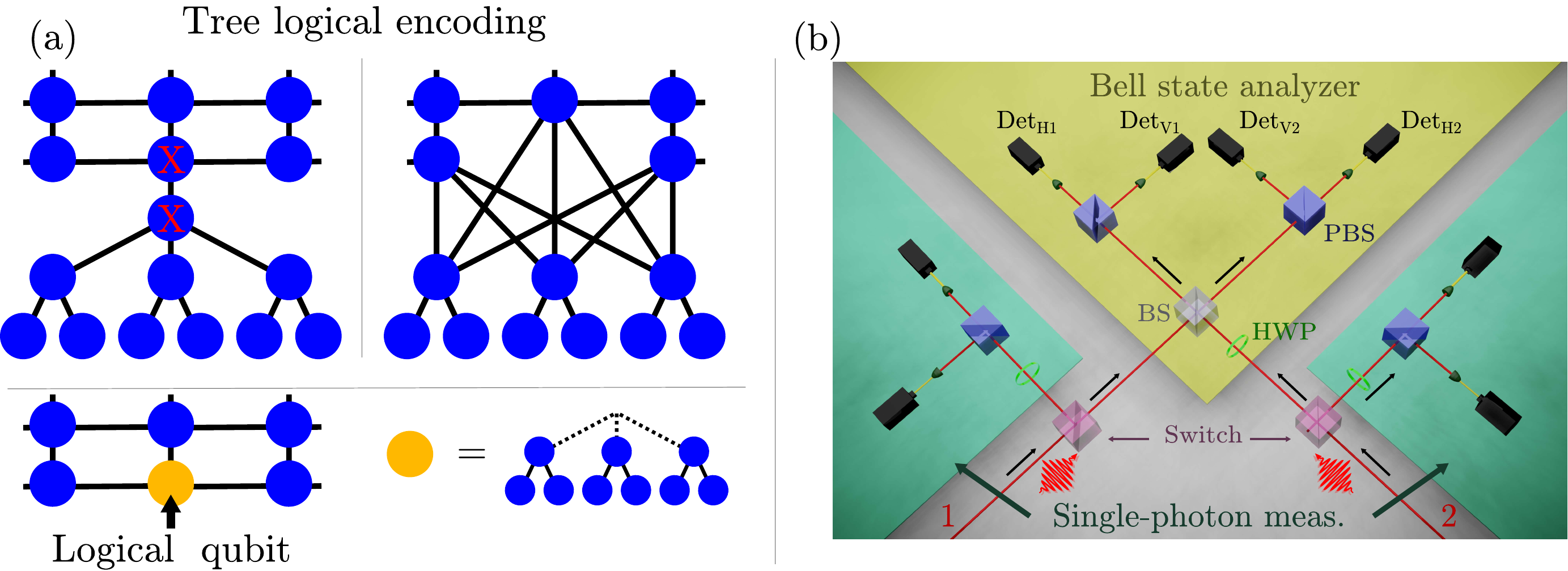}
  \caption{(a) Tree encoding of a qubit: (Left panel) The target qubit is attached to a tree via a $CZ$ gate and then $X$ measurements are performed on the target qubit and the root qubit. (Middle panel) The resulting graph and (right panel) graphical notation used for a tree-encoded logical qubit. (b) Optical setup that actively switches between a two-photon BSM, which measures $ZZ'$ unambiguously and $XX'$ partially, and two single-photon $X$ or $Z$ measurements. The two photons arrive in inputs 1 and 2. BS: beam splitter, PBS: polarization BS, HWP: half wave plate (rotated by $45^{\circ}$) and Det are single-photon detectors.}
  \label{fig_setup}
\end{figure}

We use the first recursion to explicit that the tree graph state described by $\vec{b}$, is a root qubit $0$ attached to $b_0$ tree graph states of branching vector $\vec{b}_1$ by $CZ$ gate:
\begin{equation}
  \begin{aligned}
    \ket{T_{\vec{b}}} & = \ket{0}_0 \otimes \ket{T_{\vec{b}_1}}^{\otimes b_0} + \ket{1}_0 \otimes \ket{\bar{T}_{\vec{b}_1}}^{\otimes b_0}, \\
    & = \ket{0}_0 \otimes \ket{T} + \ket{1}_0 \otimes \ket{\bar{T}}, \\
  \end{aligned}
\end{equation}
where we have used the simplifying notation $\ket{T} =\ket{T_{\vec{b}_1}}^{\otimes b_0}$.
Using the construction method presented in Fig.~\ref{fig_setup}, we logically encode a physical qubit state $\ket{\phi}_p = \alpha \ket{0}_p + \beta \ket{1}_p = (\alpha + \beta)\ket{+}_p + (\alpha - \beta) \ket{-}_p $  onto a graph state $\ket{T_{\vec{b}}}$ by performing a $CZ$ gate onto the physical qubit and the root qubit of the tree and by measuring these two qubits in the $X$ basis. After the $CZ$ gate operation, we obtain the state
\begin{equation}
  \begin{aligned}
    CZ \ket{T_{\vec{b}}} \otimes \ket{\phi}_p & = \ket{0}_0 \otimes \ket{T} \left((\alpha + \beta)\ket{+}_p + (\alpha - \beta) \ket{-}_p\right) \\
    & + \ket{1}_0 \otimes \ket{\bar{T}} \left((\alpha + \beta)\ket{-}_p + (\alpha - \beta) \ket{+}_p\right). \\
  \end{aligned}
\end{equation}

After the measurement of the root "$0$" and physiscal "$p$" qubits in the $\ket{\pm}$ basis, we obtain the logical qubit $\ket{\phi}_L$:

\begin{equation}
    \begin{aligned}
      \ket{+}_p, \ket{+}_0 & \rightarrow  \ket{\phi}_L = (\alpha + \beta) \ket{T}
      + (\alpha - \beta)  \ket{\bar{T}} , \\
    \end{aligned}
\end{equation}
For different outcomes, we can recover the same state by applying first $X_L$ if the root qubit measurement outcome is $\ket{-}_0$ and then $Z_L$ if $\ket{-}_p$;
with $Z_L$, $X_L$ the logical operations as described in the main text.

The logical state, encoded onto the tree graph state is $\ket{\phi}_L$. It follows that:
\begin{equation}
  \begin{aligned}
    & \ket{0}_L = \frac{1}{\sqrt{2}} \left(\ket{T} + \ket{\bar{T}}\right)  \; &;& \;  \ket{1}_L = \frac{1}{\sqrt{2}} \left(\ket{T} - \ket{\bar{T}}\right) \\
    & \ket{+}_L = \ket{T} \; &;&  \; \ket{-}_L = \ket{\bar{T}}, \\
    & {\rm with \;} \ket{T} = \ket{T_{\vec{b}_1}}^{\otimes b_0}. & &
  \end{aligned}
\end{equation}

Note that this logical encoding is compatible with indirect $Z$ measurements of qubits, for qubits at the first level and deeper. Indeed, any $K_v$ for qubit $v$ at the second level of the initial tree or deeper stabilize both $\ket{T}$ and $\ket{\bar{T}}$, and thus any logical state $\ket{\phi}_L$.

An alternative description of a logical code (which encode one logical qubit) is through its logical operators $Z_L$, $X_L$, and its $n-1$ stabilizers, where $n$ is the number of physical qubits of the code. As stated before, for any physical qubit $v$ at level-$2$ or deeper in the tree, the graph stabilizer $K_v$ is also a stabilizer of the code. Therefore, there remains $b_0 -1$ stabilizers to find. Since the $X_L$ operator can be applied using each of the $b_0$ first-level qubits of the tree, we can rewrite it by denoting which qubit we are using to apply this operator:
\begin{equation}
    X_{L,v} = X_v \prod_{w \in \mathcal{N}_v} Z_w
\end{equation}.
From this, we can find the remaining $b_0 - 1$ independent stabilizers of the code, which are given by the products $X_{L,i} X_{L,j}$ for any pair of first-level qubits, $i$ and $j$.

\section{Replacing a logical measurement by physical measurements}
\label{app_counterfactual}
In this section, we discuss why the measurement of logical operators such as $X_L$ or $Z_L^{ } Z_L'$, which are multi-qubit operators, can be replaced by many single-qubit or two-qubit measurements. We begin by  illustrating the problem with a minimal example using two simple trees with branching parameters $\vec b = (2)$ (depth-one tree with only three physical qubits). In that case, $\ket{T} = \ket{++}$.
An arbitrary product states of two logical qubits is:
\begin{equation}
  \begin{aligned}
    \ket{\psi}_L  \ket{\psi'}_L & = \mu \mu' \ket{++++} + \mu \nu' \ket{++--} \\ &  + \nu \mu' \ket{--++} + \nu \nu \ket{----}
  \end{aligned}
\end{equation}

We need to do physical Bell measurements on pairs of qubits taken from $\ket{\psi_L}$ and  $\ket{\psi'_L}$, and so to facilitate this, we switch the order of qubits 2 and 3:

\begin{widetext}
\begin{equation}
  \begin{aligned}
    \ket{\psi}_L  \ket{\psi'}_L&  =  \mu \mu' \ket{++++} + \mu \nu' \ket{+-+-} + \nu \mu' \ket{-+-+} + \nu \nu \ket{----} \\
    & = (\mu \mu' + \nu \nu') (\ket{\phi^+}\ket{\phi^+} + \ket{\psi^+}\ket{\psi^+})  + (\mu \mu' - \nu \nu') (\ket{\phi^+}\ket{\psi^+} + \ket{\psi^+}\ket{\phi^+}) \\
    & + (\mu \nu' + \nu \mu') (\ket{\phi^-}\ket{\phi^-} + \ket{\psi^-}\ket{\psi^-})  + (\nu \mu' - \mu \nu') (\ket{\phi^-}\ket{\psi^-} + \ket{\psi^-}\ket{\phi^-}). \\
  \end{aligned}
  \label{eq_before_ex}
\end{equation}
\end{widetext}

In this case, $Z{ }_L Z'_L = Z_1 Z_2 Z_3 Z_4$ and we have two options for $X^{ }_L X'_L$: $X^{ }_L X_L' = X_1 X_2$
or $X^{ }_L X_L' = X_3 X_4$. In the static protocol, we reconstruct the outcomes for these logical measurements from measurements of $Z_1 Z_2$, $X_1 X_2$, $Z_3 Z_4$, $X_3 X_4$, which are obtained from physical Bell measurements. An ordinary linear-optical Bell measurement on two photons only successfully yields the outcome for both $ZZ'$ and $XX'$ with probability $1/2$. For concreteness, we can account for this by assuming that the $XX'$ measurement only succeeds if $ZZ'$ yields $+1$.
As an example, suppose the physical Bell measurements yield the following outcomes: $Z_1 Z_2 \rightarrow +1$, $X_1 X_2 \rightarrow +1$, $Z_3 Z_4 \rightarrow -1$, $X_3 X_4 \rightarrow$ failed. The logical two-qubit state collapses:
\begin{equation}
  \ket{\psi_L}  \ket{\psi'_L} \rightarrow \ket{\phi^+} \ket{\psi^+}.
  \label{eq_result_ex}
\end{equation}
On the other hand, the outcomes of the logical operators in this case are $Z^{ }_L Z'_L \rightarrow -1$ and $X^{ }_L X_L' \rightarrow +1$, and the corresponding logical Bell state is
\begin{equation}
  \ket{\psi^+}_L = \ket{TT} - \ket{\bar{T} \bar{T}} = \ket{++++} - \ket{----},
\end{equation}
which remains the same if we switch the ordering of qubits 2 and 3. Obviously, this
is not the state we obtained above in Eq.~\eqref{eq_result_ex}. A similar finding occurs regardless of
what the outcomes of the physical measurements are. This can be seen from the fact
that every term in Eq.~\eqref{eq_before_ex} combines two Bell states with the same $XX'$ eigenvalue,
so that only one of these eigenvalues is needed to uniquely identify a single term
in this state. Thus, so long as one of the physical Bell measurements succeeds, the
state collapses to a single term, and it does not get projected onto a logical Bell
state. The only way to project onto a logical Bell state is to measure $Z_1 Z_2 Z_3 Z_4$
without separately measuring $Z_1 Z_2$ and $Z_3 Z_4$ .
If we performed a true $Z_1 Z_2 Z_3 Z_4$
measurement and obtained outcome $-1$ (and also obtained $X^{ }_L X'_L \rightarrow +1$), then the
state would instead collapse to
\begin{equation}
  \ket{\psi}_L \ket{\psi'}_L \rightarrow \ket{\phi^+} \ket{\psi^+} + \ket{\psi^+} \ket{\phi^+} = \ket{++++} - \ket{----},
\end{equation}
which is the desired logical Bell state. Thus, if the goal is to project the system onto
a logical Bell state, physical Bell measurements on pairs of qubits do not suffice.

However, in practice, the photonic qubits are generally absorbed by the photon detectors and the question of the outcome state becomes irrelevant since it cannot be used again after a measurement. Therefore, we consider that a set of single-qubit or two-qubit measurements is a measurement of a logical operator if the following conditions are met. The set of measurements should have the same outcome probabilities as the logical operator measurements, and
the total state after the set of measurements should be a product state $\ket{\Psi_{\rm out}} \otimes \ket{\Psi_{\rm qubit}}$ where $\ket{\Psi_{\rm qubit}}$ corresponds to the measured qubit (or qubits) subspace and $\ket{\Psi_{\rm out}}$ corresponds to the state of the other qubits that were not measured. After the set of single- or two-qubit measurement, the state $\ket{\Psi_{\rm out}}$ should be the one expected by the measurement of the logical operator. In other words, a set of many single-qubit and two-qubit measurements is a measurement of a logical operator if it acts the same way on the qubits outside of the logical subspace.

To see that this is the case here, we keep track of the qubit subpsace state while progressively realizing the measurements of the protocol. When all but one qubit in the operator is correctly measured, we see that the measurement of the logical operator is mapped onto this last physical qubit measurement. Let's take $Z_L$ and a general quantum state $\ket{0}_L \ket{\Psi_{\rm out}} + \ket{1}_L \ket{\bar{\Psi}_{\rm out}}$, as an illustration. This can be easily generalized to two-qubit measurements.
The qubit subspace is described by the logical operators and by the stabilizers:
\begin{equation}
  \begin{aligned}
    Z_L & = \prod_{v \in \mathcal{C}_0} Z_v, \\
    X_L & = X_v \prod_{w \in \mathcal{C}_v} Z_w,  \forall v \in \mathcal{C}_0, \\
    K_u & = X_u \prod_{r \in \mathcal{N}_u} Z_r,  \forall u \in V \backslash \mathcal{C}_0,
  \end{aligned}
  \label{eq_stabs}
\end{equation}
where the stabilizers are for all the vertices of the tree, except the first-level qubits.

After one qubit measurement, say $Z_v$ with outcome $m = \pm 1$, we can recover the new qubit subspace by following these rules:
\begin{itemize}
  \item For all the operators $S$ that act on qubit $v$ trivially (with identity operator $I_v$), $S$ is not modified by the measurement.
  \item For all $S$ containing $Z_v$ (i.e. the measurement basis), $S$ is converted into $S' = m Z_v S$ (similar to $S$ except that $Z_v$ is replaced by $m I_v$).
  \item The remaining operators contain $X_v$ (if there are operators containing $Y_v$, we can multiply them by a previous operator containing $Z_v$). If there is only one such operator, we can replace it by $ m Z_v$, stating that after the measurement the qubit is in state $\ket{0}_v$ or $\ket{1}_v$.  Considering destructive measurements, we can also discard this operator since the measured photon does not exist anymore. If there is more than one operator containing $X_v$, we denote them by $S_0$, $S_1$, ... $S_N$. $S_0$, $S_0S_1$, $S_0S_2$, ..., $S_0S_N$ is another set of independent stabilizers where only $S_0$ contains $X_v$ (the others contain $I_v$). After the measurement we replace $S_0$ by $m Z_v$ and we keep the other one containing $I_v$.
\end{itemize}

We illustrate this with a three-qubit linear cluster state: $\ket{\psi} = \ket{+0+} + \ket{-1-}$, stabilized by $\{ X_1 Z_2 I_3, Z_1 X_2 Z_3, I_1 Z_2 X_3 \}$. If we measure the second qubit in $Z_2$ and apply these rules we obtain as expected the stabilizers $\{ mX_1 I_2 I_3 , mI_1 Z_2 I_3, mI_1 I_2 X_3 \}$ corresponding to the states $\ket{\psi_{m=1}} = \ket{+0+}$ or  $\ket{\psi_{m=-1}} = \ket{-1-}$.
If  we perform an $X_2$ measurement on the second qubit instead of a $Z_2$ measurement, the stabilizer set is $\{ mZ_1 I_2 Z_3, X_1 I_2 X_3, mI_1 X_2 I_3 \}$ which corresponds to $\ket{\psi_{m=1}} = \ket{0+0} + \ket{1+1}$ or  $\ket{\psi_{m=-1}} = \ket{0-1}+  \ket{1-0}$.

Going back to the calculation of $Z_L$ on a tree (see Eq.~\eqref{eq_stabs}) and measuring all but one qubit $v'$ in the first level, we obtain:
\begin{equation}
  \begin{aligned}
    Z_L & = \left(\prod_{v \in \mathcal{C}_0 \backslash \{v'\}} m_v \right) Z_{v'}, & &\\
    X_L & = X_{v'} \prod_{w \in \mathcal{C}_{v'}} Z_w, & & \\
    K_u & = X_u \prod_{r \in \mathcal{N}_u} Z_r, &  &\forall u \in V \backslash (\mathcal{L}_1 \cup \mathcal{L}_2 ), \\
    K_u & = m_v X_u \prod_{r \in \mathcal{C}_u} Z_r, & &\forall v \in \mathcal{C}_0 \backslash v', \forall u \in \mathcal{C}_v, \\
    K_u & = Z_{v'} X_u \prod_{r \in \mathcal{C}_u} Z_r, & &\forall u \in \mathcal{C}_{v'},
  \end{aligned}
\end{equation}

where $m_v$ is the measurement outcome of $Z_v$, and we used $\mathcal{L}_k$ to denote the set of all the qubits at level $k$ (e.g. $\mathcal{L}_1 = \mathcal{C}_0$).
After all of these measurements, we observe that $Z_L$ is indeed mapped onto the measurement of the last physical qubit (up to a sign that depends on the previous measurements). Therefore, when all the other qubits were correctly measured, the effect of this last $Z_{v'}$ physical measurement, yields the same effect as $Z_L$ on the remaining qubits outside of the logical qubit subspace.

\section{Error analysis}
\label{app_error}

\subsection{Error of a two-photon BSM}
The following error and performance analyses depends on the type of two-photon linear optical setup. We use the one presented in Fig.~\ref{fig_setup}(b), which allows to actively switch between single-photon measurements and a two-photon BSM that measure $ZZ'$ unambiguously and $XX'$ only when $ZZ' \rightarrow 1$. Note that for the static protocol only requires the central part of this setup (in yellow), and the active switches and single-photon measurements can be removed.

We assume that the photons in each tree have a single-qubit depolarization error rate $\varepsilon$, i.e. $\mathcal{E}[\mathcal{D}_{W}] = \varepsilon$, for $W=X$ or $Z$.
It corresponds to a depolarization channel:
\begin{equation}
  \begin{aligned}
    \mathcal{E}(\rho) = (1 - \varepsilon_{d}) \rho + \frac{\varepsilon_{d}}{3}  ( X \rho X + Y \rho Y  +  Z \rho Z ),\\
  \end{aligned}
\end{equation}
with $\varepsilon_{d} = \frac{3}{2} \varepsilon$.

We now derive the Bell state measurement error for two qubits from these two trees.

The error induced on the density matrix that characterizes the two qubits is therefore:

\begin{equation}
  \begin{aligned}
    \mathcal{E} \circ \mathcal{E}'(\rho) & = (1 - \varepsilon_{d}) (1 - \varepsilon_{d}) \rho\\
    & + (1 - \varepsilon_{d}) \frac{\varepsilon_{d}}{3} (X \rho X + Y \rho Y + Z \rho Z) \\
    & + (1 - \varepsilon_{d}) \frac{\varepsilon_{d}}{3} (X' \rho X' + Y' \rho Y' + Z' \rho Z') \\
    & + \frac{{\varepsilon_{d}}^2}{9} \sum_{\substack{W \in \{X, Y, Z\} \\ W' \in \{X', Y', Z'\}  }} W W' \rho W' W, \\
  \end{aligned}
\end{equation}
where we use the prime to distinguish operators acting on the second qubit.

To understand better the effect of the depolarization on a BSM, we summarize here the effect of the Pauli operators on the stabilizers $ZZ'$ and $XX'$:
\begin{equation}
  \begin{aligned}
    X (X X') X & = X X'; & X (Z Z') X & = -Z Z', \\
    Y (X X') Y & = -X X'; & Y (Z Z') Y & = -Z Z', \\
    Z (X X') Z & = -X X'; & Z (Z Z') Z & = Z Z', \\
  \end{aligned}
\end{equation}
and similarly for $X',Y',Z'$.

For these different terms, if only one Pauli matrix is applied, this leads to at least one error in the stabilizer measurement ($ZZ'$ or $XX'$) of the Bell state measurement. The error on the two qubits can compensate themselves only if the Pauli matrices applied are the same for the two qubits (i.e. for $X X' \rho X' X$, $Y Y' \rho Y' Y$ or $Z Z' \rho Z' Z$).
The term in $\mathcal{E} \circ \mathcal{E}'(\rho)$ that is errorless is therefore $((1 - \varepsilon_{d}) (1 - \varepsilon_{d}) + \frac{\varepsilon_{d}^2}{3}) \rho$, corresponding to an error rate of:
\begin{equation}
  \varepsilon_{\rm BSM} = 2\varepsilon_{d} - \frac{4{\varepsilon_{d}}^2}{3} = 3 \varepsilon (1 - \varepsilon)
  \label{eq_err_bsm}
\end{equation}

For the $XX'$ measurement of the BSM only, due to the way these measurements are realized, both the $ZZ'$ and the $XX'$ measurement should succeed, as otherwise it leads to an indeterminate result, so:

\begin{equation}
  \mathcal{E}[\mathcal{D}_{XX}] = \varepsilon_{\rm BSM}.
  \label{eq_err_xx_bsm}
\end{equation}

For the $ZZ'$ measurement, however, an error on the $XX'$ parity measurement does not lead to an error on the $ZZ'$ measurement. So the error should be smaller than $\varepsilon_{\rm BSM}$.
If there is only one single-qubit error, (a single Pauli matrix applied on $\rho$), this does not lead to errors if this Pauli matrix is either $Z$ or $Z'$.
For errors applied on two qubits, as in the first case, there is no error if the Pauli matrices applied on the two qubits are the same, but also if $X Y'$ or $Y X'$ matrix are applied. The error-less term in that case is therefore:
\begin{equation}
  \begin{aligned}
    \mathcal{E}[\mathcal{D}_{ZZ}] & = \varepsilon_{\rm BSM} - 2(1 - \varepsilon_{d}) \frac{\varepsilon_{d}}{3} - \frac{2 {\varepsilon_{d}}^2}{9} \\
    & = \frac{2}{3} \varepsilon_{\rm BSM}.
  \end{aligned}
  \label{eq_err_zz_bsm}
\end{equation}

Similarly to Eqs~\ref{eq_pr_m_bsm_l}, the error probability for the complete logical BSM is:
\begin{equation}
    \mathcal{E}[\mathcal{M}_{{\rm BSM}, L}^{(c)}] = \mathcal{E}[{\mathcal{M}_{Z_L Z_L'}}] + (1 - \mathcal{E}[{\mathcal{M}_{Z_L Z_L'}}]) \mathcal{E}[{\mathcal{M}_{X_L X_L'}}].
\end{equation}

\subsection{Logical qubit error analysis}

We denote by $\mathcal{E}[A]$, the error probability of the event $A$.
The error correction is based on the fact that the error of indirect measurements can be reduced thanks to a majority vote on indirect measurements.

Therefore, for the error correction to work, the error of an indirect measurement must be lower than that of a direct measurement $\mathcal{E}[\mathcal{I}_{W,k}] \leq \mathcal{E}[\mathcal{D}_{W,k}]$. Consequently, we should rely preferrably on the indirect measurement outcomes:
\begin{equation}
  \begin{aligned}
    \mathcal{E}[\mathcal{M}_{W,k}] & = \mathrm{Pr}[\mathcal{I}_{W,k}|\mathcal{M}_{W,k}] \mathcal{E}[I_{W,k}] \\ & + (1 - \mathrm{Pr}[\mathcal{I}_{W,k}|\mathcal{M}_{W,k}]) \mathcal{E}[\mathcal{D}_{W,k}]
  \end{aligned}
    \label{eq_err_m_wk}
\end{equation}
where $\mathrm{Pr}[A|B]$ denotes the conditional probability of $A$ given $B$:
\begin{equation}
    \mathrm{Pr}[\mathcal{I}_{W,k}|\mathcal{M}_{W,k}]  = \frac{\mathrm{Pr}[\mathcal{I}_{W,k}]}{\mathrm{Pr}[\mathcal{M}_{W,k}]},
    \label{eq_pr_i_wk_cond_m_wk}
\end{equation}
since $\mathrm{Pr}[\mathcal{M}_{W,k}|\mathcal{I}_{W,k}] = 1$.

The qubits situated at the last level $k=d$ can only be directly measured:
\begin{equation}
  \mathcal{E}[\mathcal{M}_{W,d}] = \mathcal{E}[\mathcal{D}_{W,d}]
\end{equation}

In addition, the error of an indirect measurement $\mathcal{I}_{W,k}$ is given by:
\begin{equation}
    \mathcal{E}[\mathcal{I}_{W,k}] = \frac{1}{\mathrm{Pr}[\mathcal{I}_{W,k}]} \sum_{m_s=1}^{b_k} \mathrm{Pr}[\mathcal{I}_{W,k}, m_s] \mathcal{E}[\mathcal{I}_{W,k}, m_s],
    \label{eq_err_i_wk}
\end{equation}
where $\mathrm{Pr}[\mathcal{I}_{W,k}, m_s]$ denotes the probability of having $m_s$ individual indirect measurements $\mathcal{S}_{W,k}$ that have succeeded:
\begin{equation}
    \mathrm{Pr}[\mathcal{I}_{W,k}, m_s] = \binom{b_k}{m_s} \mathrm{Pr}[\mathcal{S}_{W,k}]^{m_s} (1 - \mathrm{Pr}[\mathcal{S}_{W,k}])^{b_k - m_s}
    \label{eq_pr_i_wk_ms}
\end{equation}
and $\mathcal{E}[\mathcal{I}_{W,k}, m_s]$ is the error probability for $m_s$ indirect individual measurements. We use a majority vote to reduce this error.
Given that $m_s$ indirect measurements are successful, an error still occurs in the majority vote if more than half of the indirect measurements ($m_s/2$) are faulty:

\begin{widetext}
  \begin{equation}
      \begin{aligned}
          \mathcal{E}[\mathcal{I}_{W,k}, m_s] & = \sum_{i = \lceil m_s/2 \rceil}^{m_s} \binom{m_s}{i} \mathcal{E}[\mathcal{S}_{W,k}]^{i} (1 - \mathcal{E}[\mathcal{S}_{W,k}])^{m_s - i}, & m_s \; \mathrm{odd} \\
           & = \sum_{i = m_s/2 }^{m_s-1} \binom{m_s-1}{i} \mathcal{E}[\mathcal{S}_{W,k}]^{i} (1 - \mathcal{E}[\mathcal{S}_{W,k}])^{m_s -1 - i}, & m_s \; \mathrm{even}
      \end{aligned}
      \label{eq_err_i_wk_ms}
  \end{equation}

For the even case, the sum goes only up to $m_k - 1$ because we cannot do better than randomly removing one result and return to the odd case.

  \begin{equation}
      \mathcal{E}[\mathcal{S}_{W,k}] = \sum_{i=0}^1 \mathcal{E}[\mathcal{D}_{\widetilde{W},k+1}]^{i} (1 - \mathcal{E}[\mathcal{D}_{\widetilde{W},k+1}])^{1 - i}
      \sum_{\substack{j = 0,\\ i+j {\rm \; odd}}}^{b_{k+1}} \binom{b_{k+1}}{j}
         \mathcal{E}[\mathcal{M}_{W,k+2}]^j (1 - \mathcal{E}[\mathcal{M}_{W,k+2}])^{b_{k+1} - j}
      \label{eq_err_s_wk}.
  \end{equation}
\end{widetext}

In this equation, we only consider odd numbers of errors because in a parity measurement even numbers of errors compensate each other.

For logical measurements, we therefore have:
\begin{align}
  \mathcal{E}[\mathcal{M}_{X_L X_L'}] & = \mathcal{E}[\mathcal{I}_{ZZ', 0}] \\
  \mathcal{E}[\mathcal{M}_{Z_L Z_L'}] & = \sum_{\substack{i=1 \\ i = 1 [2]}}^{b_0} \binom{b_0}{i} \mathcal{E}[\mathcal{M}_{ZZ', 1}]^i  (1 - \mathcal{E}[\mathcal{M}_{ZZ', 1}])^{b_0 - i},
\end{align}
where the index $i$ takes odd values ($i=1[2]$), since even numbers of parity errors lead to a correct global parity measurement outcome.

\section{Dynamic protocol}
\label{app_dynamic}

In the dynamic protocol, the type of measurements performed depends on the measurement outcome of the parent qubits. We need to discriminate the three BSM outcomes: complete ($c$), partial ($p$) and failed ($f$). Now the $ZZ'$ measurement probabilities at level $k$ are given by:
\begin{equation}
  \mathrm{Pr}[\mathcal{M}_{ZZ',k}] = \eta^2+ (1 - \eta^2) \mathrm{Pr}[\mathcal{I}_{ZZ',k}, f].
\end{equation}
Indeed, if the measurement is complete or partial (with probability $\eta^2$), $ZZ'$ is measured but if the measurement has failed ($f$), we should indirectly measure it with probability $\mathrm{Pr}[\mathcal{I}_{ZZ',k}, f]$.
If a two-photon BSM fails ($f$) or is partial ($p$), it is impossible to recover indirectly the $XX'$ components.
But in that case, the indirect $ZZ'$ measurement at level $k$ can also be performed with higher probability via two single-qubit measurements, which have success probabilities $\mathrm{Pr}[\mathcal{I}_{Z,k}]$ and $\mathrm{Pr}[\mathcal{I}_{Z',k}]$ and errors $\mathcal{E}[\mathcal{I}_{Z,k}]$ and  $\mathcal{E}[\mathcal{I}_{Z',k}]$, respectively. Therefore,
\begin{equation}
  \begin{aligned}
    \mathrm{Pr}[\mathcal{I}_{ZZ', k}, f]  & = \mathrm{Pr}[\mathcal{I}_{ZZ', k}, p] = \mathrm{Pr}[\mathcal{I}_{Z, k}] \mathrm{Pr}[\mathcal{I'}_{Z, k}], \\
    \mathcal{E}[\mathcal{I}_{ZZ', k}, f] & = \mathcal{E}[\mathcal{I}_{ZZ', k}, p] \\
    & = \mathcal{E}[\mathcal{I}_{Z, k}] (1 - \mathcal{E}[\mathcal{I}_{Z', k}]) +  \mathcal{E}[\mathcal{I}_{Z, k}] (1 - \mathcal{E}[\mathcal{I}_{Z', k}])\\
    & = \mathcal{E}[\mathcal{I}_{Z, k}] + \mathcal{E}[\mathcal{I}_{Z', k}] - 2 \mathcal{E}[\mathcal{I}_{Z, k}] \mathcal{E}[\mathcal{I}_{Z', k}].\\
  \end{aligned}
\end{equation}
Here again, we consider that a combination of two errors would yield the correct outcome. For a complete measurement, the indirect $ZZ'$ measurement probability is again given by:
\begin{equation}
  \mathrm{Pr}[\mathcal{I}_{ZZ', k}, c] = 1 - (1 - \mathrm{Pr}[\mathcal{S}_{ZZ', k}])^{b_k}
\end{equation}.

The error analysis in the case of a complete measurement is more complicated since we need to keep track of all the measurement probabilities at each level.

Because we are performing BSMs on the child qubits, once again we have three different outcomes: complete (both $XX'$ and $ZZ'$ are measured), partial (only $ZZ'$) or failed (no outcome).
For a successful BSM on child qubits denoted $B$ and $B'$ at level $k+1$, BSMs are also performed on all the pairs of child qubits of $B$ and $B'$. An indirect measurement of the qubit at level $k$ thus requires that all the children of $B$ are measured at least in $ZZ'$. An individual indirect measurement of $ZZ'$ at level $k$, $\mathcal{S}_{ZZ',k}$, in this setting thus requires the successful BSM of $B$ and $B'$ (with probability $P_{\rm Bell}$) and the measurements of $ZZ'$ on all the child qubits of $B$ and $B'$. Let's denote by $m^{(c)}$, $m^{(p)}$ and $m^{(f)}$ the number of complete, partial, and failed BSMs performed on the child qubits of $B$ and $B'$.
A successful indirect measurement of $ZZ'$ (with or without errors) occurs with probability:
\begin{widetext}
\begin{equation}
    \mathrm{Pr}[\mathcal{S}_{ZZ', k},c] = \frac{\eta^2}{2} \sum_{\substack{m^{(c)} + m^{(p)} + m^{(f)} = b_{k+1} \\ m^{(c)}, m^{(p)}, m^{(f)} \geq 0}} P_{\rm BSM}(m^{(c)}, m^{(p)}, m^{(f)}) (\mathrm{Pr}[\mathcal{I}_{Z,k+2}]\mathrm{Pr}[\mathcal{I}_{Z',k+2}])^{m^{(f)}}
\end{equation}
\end{widetext}
Here $\eta^2/ 2$ signifies that the BSM on $B$ and $B'$ has to succeed, the sum goes through all the possible BSM outcomes on the child qubits, and $(\mathrm{Pr}[\mathcal{I}_{Z,k+2}]\mathrm{Pr}[\mathcal{I'}_{Z,k+2}])^{m^{(f)}}$ accounts for the fact that when the BSMs fail, these child qubits should be indirectly measured.

The error of an individual indirect measurement with $m^{(c)}$ complete and $m^{(p)}$ partial measurements is given by:
\begin{widetext}
  \begin{equation}
    \begin{aligned}
        \mathcal{E}[\mathcal{S}_{ZZ',k}|m^{(c)}, m^{(p)}, m^{(f)}, c] =
        \sum_{i=0}^{1} \sum_{j=0}^{m^{(c)}} \sum_{k = 0}^{m^{(p)}}& \sum_{\substack{l = 0,\\ i+j+k+l=1[2]}}^{m^{(f)}}
          {\varepsilon_{\rm Bell}}^i (1 - \varepsilon_{\rm Bell})^{1 -i}  \\
         \times &  \binom{m^{(c)}}{j} {\mathcal{E}[\mathcal{M}_{ZZ',k+2}, c]}^{j} (1 - \mathcal{E}[\mathcal{M}_{ZZ',k+2}, c])^{m^{(c)} - j} \\
         \times & \binom{m^{(p)}}{k}  {\mathcal{E}[\mathcal{M}_{ZZ',k+2}, p]}^k (1 - \mathcal{E}[\mathcal{M}_{ZZ',k+2}, p])^{m^{(p)} - k} \\
         \times & \binom{m^{(f)}}{l} {\mathcal{E}[\mathcal{M}_{ZZ',k+2}, f]}^l (1- \mathcal{E}[\mathcal{M}_{ZZ',k+2}, f])^{m^{(f)} - l}.
    \end{aligned}
\end{equation}

The individual indirect error probability is therefore:
\begin{equation}
    \mathcal{E}[S_{ZZ',k}, c] = \sum_{\substack{m^{(c)} + m^{(p)} + m^{(f)} =b_{k+1} \\ m^{(c)}, m^{(p)}, m^{(f)} \geq 0}} P_{\rm BSM}(m^{(c)}, m^{(p)}, m^{(f)}) \mathcal{E}[\mathcal{S}_{ZZ', k}|m^{(c)}, m^{(p)}, m^{(f)}, c].
\end{equation}

If $m_k$ indirect measurements are performed, the error is therefore:
\begin{equation}
    \begin{aligned}
        \mathcal{E}[\mathcal{I}_{ZZ',k}|m_k, c] & = \sum_{i=\lceil m_k/2 \rceil}^{m_k} \binom{m_k}{i} {\mathcal{E}[\mathcal{S}_{ZZ',k}, c]}^i  (1 - \mathcal{E}[\mathcal{S}_{ZZ',k}, c])^{m_k - i}, & m_k \mathrm{\; odd},   \\
        & = \sum_{i=m_k/2}^{m_k-1} \binom{m_k-1}{i} {\mathcal{E}[\mathcal{S}_{ZZ',k},c]}^i  (1 - \mathcal{E}[\mathcal{S}_{ZZ',k},c])^{m_k - 1 - i}, &  m_k \mathrm{\; even}.   \\
    \end{aligned}
\end{equation}
\end{widetext}

Finally, we find the error probability of an indirect measurement to be:
\begin{equation}
  \begin{aligned}
    &\mathcal{E}[\mathcal{I}_{ZZ',k}, c] \\ & = \frac{1}{\mathrm{Pr}[\mathcal{I}_{ZZ', k},c]}\sum_{m_k=1}^{b_k} \mathrm{Pr}[\mathcal{I}_{ZZ',k},m_k,c] \mathcal{E}[\mathcal{I}_{ZZ',k}|m_k, c],
  \end{aligned}
\end{equation}
with
\begin{equation}
  \begin{aligned}
    & \mathrm{Pr}[\mathcal{I}_{ZZ',k},m_k, c] \\ & = \binom{b_k}{m_k} {\mathrm{Pr}[\mathcal{S}_{ZZ', k},c]}^{m_k} {(1 - \mathrm{Pr}[\mathcal{S}_{ZZ', k},c])}^{b_k - m_k},
  \end{aligned}
\end{equation}
and
\begin{equation}
  \begin{aligned}
    & \mathrm{Pr}[\mathcal{I}_{ZZ', k},c] \\ & = \sum_{m_k=1}^{b_k} \mathrm{Pr}[\mathcal{I}_{ZZ', k},m_k, c] = 1 - (1 - \mathrm{Pr}[\mathcal{S}_{ZZ', k}, c])^{b_k}.
  \end{aligned}
\end{equation}

\section{``Loss-only'' adaptive protocol}
It is also worth noting that in principle, it is also possible to realize a loss-tolerant BSM by performing single-qubit measurements on all qubits below level $1$, but this strategy fails to enable error correction. Indeed, in that case, the child qubits of a complete BSM ($Z_v Z_{v'}$ and $X_v X_{v'}$) at level $1$ need to be measured in the $Z$ basis to measure $X^{ }_L X'_L$ so that they cannot provide an indirect $Z_v Z_{v'}$ measurement of the qubit at level $1$, which is necessary for error correction of $Z^{ }_L Z'_L$. Error correction is therefore impossible with this protocol and the improvement of loss-tolerance is relatively small compared to the dynamic protocol.

\bibliography{Biblio}

\end{document}